\documentclass[twocolumn]{aastex631}

\shorttitle{Coronal hole fast wind}
\shortauthors{Bandyopadhyay et al.}

\begin{document}

\title{First Observation of a Polar Coronal Hole-like Fast Solar Wind Stream in the Sub-Alfv\'enic Solar Corona: an Analysis of Turbulence Properties}

\author{Riddhi Bandyopadhyay}
\affiliation{Department of Physics and Astronomy, University of Delaware, Newark, DE 19716, USA}

\author{Sujan Prasad Gautam}
\affiliation{Department of Physics and Astronomy, University of Delaware, Newark, DE 19716, USA}

\author{William H. Matthaeus}
\affiliation{Department of Physics and Astronomy, University of Delaware, Newark, DE 19716, USA}

\author{Yeimy J. Rivera}
\affiliation{Smithsonian Astrophysical Observatory--Harvard \& Smithsonian, Cambridge, MA 02138, USA}

\author{Samuel T. Badman}
\affiliation{Smithsonian Astrophysical Observatory--Harvard \& Smithsonian, Cambridge, MA 02138, USA}

\begin{abstract}
Parker Solar Probe, near its $23^{\mathrm{rd}}$ perihelion in March 2025, sampled an extended interval of sub-Alfv\'enic solar wind likely originating from a large equatorial coronal hole. At heliocentric distances of approximately $10\,R_{\odot}$, with speed mostly above $400~\mathrm{km\,s^{-1}}$, this interval is a first example of ``polar coronal hole-like (PCH-l) fast" solar wind observed in the sub-Alfv\'enic solar corona. We characterize the turbulence properties of this unique interval using Parker Solar Probe measurements. Despite being sampled well inside the nominal Alfv\'en surface, the turbulence appears to be already well developed while remaining strongly transverse and highly imbalanced, exhibiting a large cross helicity. These observations provide new constraints on the development and evolution of solar wind turbulence within the lower corona.
\end{abstract}


\section{Introduction} \label{sec:intro}
A standard and highly simplified picture of the solar wind is that high speed, highly Alfv\'enic wind emerges from open field line coronal hole regions, while lower speed, less Alfv\'enic wind 
originates from
sources at 
lower latitude, generally thought to be 
closed field line regions.
This idealized view pertains mainly to solar minimum conditions, but can be extended to solar maximum phase as well. 

The pristine solar wind emerging from coronal holes has traditionally been viewed as the 
paradigm for 
coronal heating and 
solar wind acceleration models.
Models of several types
consider the fast coronal hole wind as the standard to be explained, for example
by absorption of high frequency cyclotron waves~\citep{AxfordMcKenzie-sw8, BanaszkiewiczEA98}, 
or
by reflection-driven low
frequency magetodydrodynamic (MHD) turbulence~\citep{Matthaeus1999ApJ, Cranmer2005ApJS, VerdiniEA10}. 
In concert with 
the presumptive
simple topology of the magnetic
field emanating from coronal holes, such heating and acceleration models provide a framework for 
understanding 
the monopolar fields, high wind speeds, high temperatures, low densities, and high cross helicities observed at high latitudes in solar minimum conditions,
for example by Ulysses \citep{McComasEA00}.

The orbits of Parker Solar Probe (PSP) 
remain at low heliographic latitudes
and have mainly sampled slower wind over its first $\sim 20$
orbits~\citep{Raouafi2023SSR_PSP}. Particularly, all the sub-Alfv\'enic samples observed by Parker to date have been at most marginally ``fast'' as classified by their asymptotic speed at 1 au~\citep{Rivera2024Science_heating, Adhikari2026ApJ_sub-alfvenic}  and no sub-Alfv\'enic samples have previously been made of polar coronal hole-like (PCH-l) fast wind which later reaches the typical Ulysses asymptotic speeds~\citep{McComasEA00}, and which has also been shown to be the maximum speed of high speed streams (HSS) in the ecliptic plane~\citep{Garton2018ApJL_HSS}. Here we report the observation of a ``fast 
Alfv\'enic'' wind interval sampled by PSP while the spacecraft was still in the sub-Alfv\'enic corona, and describe its plasma and turbulence properties.

\section{Observations} \label{sec:obs}
We analyze measurements obtained by the PSP spacecraft during its $23^\mathrm{rd}$ perihelion passage in March 2025, when the spacecraft reached a heliocentric distance of approximately $10\,R_{\odot}$~\citep{Guo2021ActaAstro}. During this interval, PSP encountered an extended period of fast ($V \gtrsim 400$ km s$^{-1}$) sub-Alfv\'enic solar wind. The fast sub-Alfv\'enic interval starts on March 22, 2025 and lasts until March 23, 2025. Notably, as shown in a complementary work, this wind stream was sourced from a large equatorial coronal hole which was aligned with the solar dipole axis at this time~\citep[][\& in preparation]{Rivera2025AGU_coronal-hole}, and corresponds to a high speed stream measured at 1 au which confirms it reaches the Ulysses ``saturated" speed range of 750-800 km/s~\citep{Badman2026ArXiv_polar-coronal}, and is therefore well characterized as a PCH-l fast solar wind stream. For a detailed argument associating this wind stream with a polar-like coronal hole, see \cite{Badman2026ArXiv_polar-coronal}. The precise footpoint location remains subject to uncertainties associated with ballistic propagation and coronal magnetic field extrapolation~\citep[][e.g.,]{Benavitz2024ApJ_PFSS}. These include uncertainties related to ballistic propagation, the PFSS approximation and adopted source surface height, the photospheric magnetic field map, and temporal evolution of the coronal magnetic field, all of which can shift the inferred connectivity. Nevertheless, as shown in \cite{Badman2026ArXiv_polar-coronal}, due to the longitudinal width of the stream and the size of the coronal hole, the mapping and source association are rather robust to the typical errors associated with ballistic mapping and PFSS parameter selection.

\begin{figure}
    \centering
    \includegraphics[width=\linewidth]{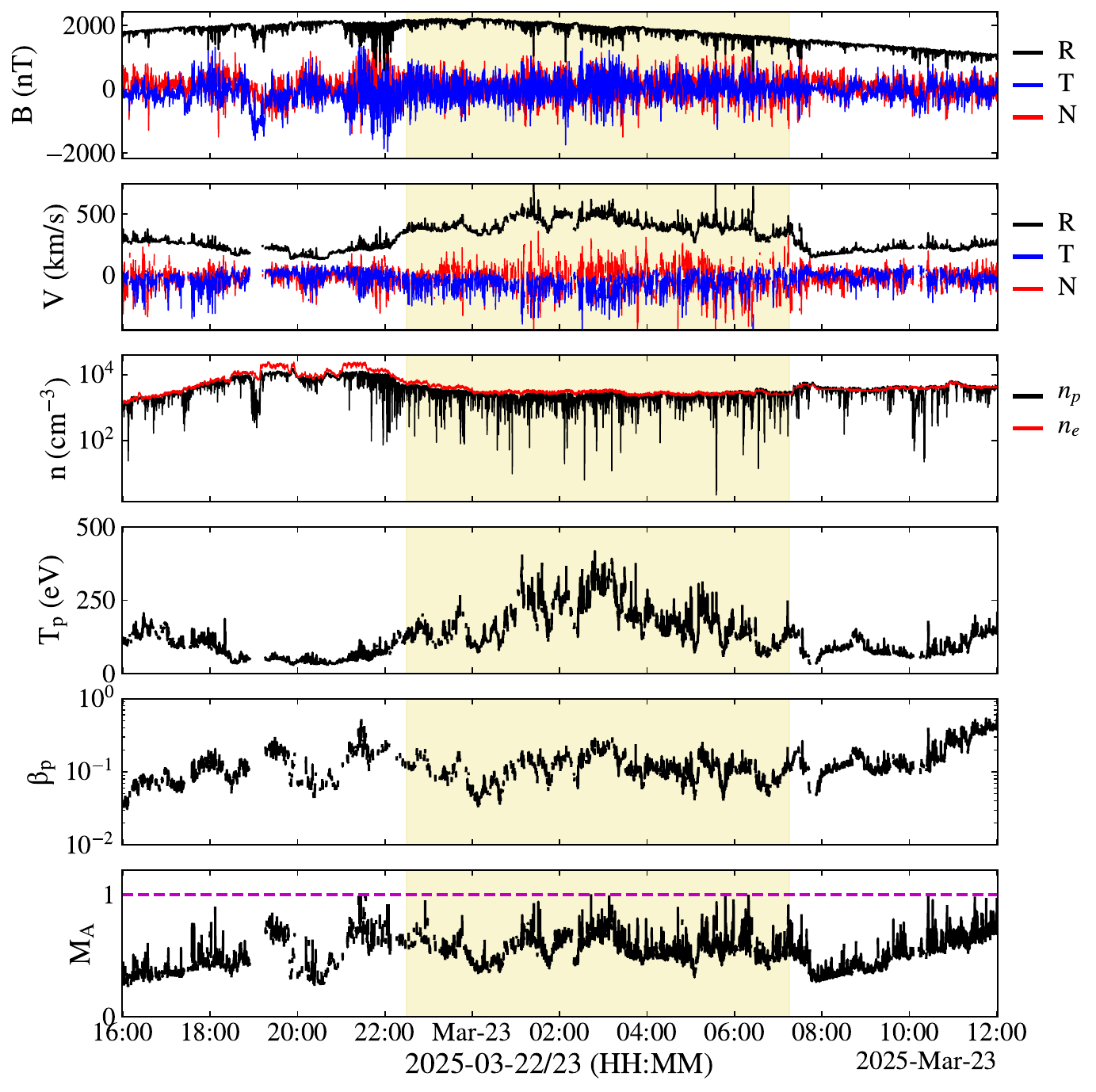}
    \caption{Overview of key plasma and magnetic field properties as a function of time during a fast sub-Alfv\'enic solar wind interval observed by PSP on 2025 March 22–23. From top to bottom, the panels show the magnetic field components ($B_R$, $B_T$, and $B_N$), solar wind velocity components ($V_R$, $V_T$, and $V_N$), proton (in black) and electron (in red) number densities ($n_p$ and $n_e$), proton temperature ($T_p$), plasma beta ($\beta_p$), and Alfv\'en Mach number ($M_A$) calculated using the QTN-based electron density. The shaded region represents the selected fast sub-Alfv\'enic interval used for the analysis.}\label{fig:overview}
\end{figure}

Figure~\ref{fig:overview} presents an overview of the plasma and magnetic field properties during the selected interval and the neighboring solar wind samples. The top panel plots the magnetic field components. The measurements are obtained from the fluxgate magnetometer (MAG) and search coil magnetometer (SCM) in the FIELDS instrument suite~\citep{Bale2016SSR, Jannet2021JGR_SCM}. Plasma moments are taken from the SWEAP/SPAN-I observations~\citep{Kasper2016SSR, Livi2022ApJ_SPANI}. In this study, we use proton plasma moments including proton bulk velocity ($\mathbf{V}$) and proton temperature ($T_p$). However, due to the proton velocity distribution function (VDF) often being outside the field of view (FOV) of SPAN-I, we use Quasi thermal noise (QTN)-based estimates~\citep{Moncuquet2020ApJS_qtn} of electron density$(n_e)$ as a proxy for the proton density $(n_p)$. 

In addition, to reduce biases associated with the limited FOV of the SPAN-I instrument, velocity fluctuations are produced only for velocity moments which are filtered to reject instances where the velocity distribution function (VDF) peak is at or beyond the FOV limit. This correction is particularly important during large +T transverse velocity deflections, where the measured instantaneous velocity moment can become a poor measurement despite the background velocity being very well observed. Velocity fluctuations are computed relative to a 10-minute averaged background field. The background velocity is estimated using the median rather than the mean because the fluctuations during this interval are largely Alfv\'enic and approximately spherical in velocity space. Furthermore, the tangential component is treated as a special case by assuming a zero large-scale tangential background $(\delta V_T = V_T)$, while the radial and normal components are computed relative to their local median values. This helps to minimize the systematic offset in the background tangential flow which arises because large $-V_T$ deflections are not strongly affected by measurement issues while $+V_T$ deflections are. This change yields a fluctuation distribution that is more consistent with the observed magnetic field fluctuations. It is worth noting that the effect of neglecting any of these effects would reduce the level of correlation between $\mathbf{V}$ and $\mathbf{B}$ and would therefore artificially reduce the apparent Alfv\'enicity of these measurements.

The limited FOV of SPAN-I can also affect the proton temperature estimate. To correct for the proton temperature, we use the SPAN-i L3 temperature tensor and follow equations A.4 and A.5 from \cite{Badman2025ApJ_alfven-surface} to reject the tensor component most strongly affected by FOV effects. We note that this increases the temperature non-negligibly in the hottest part of the streams relative to the standard L3 scalar moment (approximately $27\%$), but does not otherwise modify the overall stream structure. As noted in \cite{Badman2025ApJ_alfven-surface}, the transformation has a singularity when the magnetic field is close to the spacecraft +Y axis (which corresponds in this stream to large deflections in the tangential direction), and leads to artificial apparent spikes during large deflections which we here filter out with a two-point median filter (see Fig.~\ref{fig:overview}). The fourth panel in Fig.~\ref{fig:overview} plots the filtered proton temperature.

Plasma beta and Alfv\'en Mach number are computed using standard definitions, $\beta_p = {2 \mu_0 \, n_p \, k_B \, T_p}/{B^2}$,
and $M_A = {V}/{V_A}$, where $V_A = B/\sqrt{\mu_0 \, m_p \, n_p}$ is the Alfv\'en speed with $m_p$ being the proton mass. Note that QTN-based electron density is used for solar wind density.

From Fig.~\ref{fig:overview} the solar wind speed remains predominantly above 400 km s$^{-1}$, while the Alfv\'en Mach number ($M_A$) stays below unity for an extended period. The interval is characterized by relatively low proton density, high proton temperature, and low plasma beta, consistent with coronal-hole-associated wind~\citep{McComasEA00}. As mentioned previously, magnetic connectivity of this interval to its likely coronal-hole source region is explored in a companion study using Potential Field Source Surface (PFSS) models~\citep{Badman2026ArXiv_polar-coronal}. The proton temperature becomes substantially elevated, suggesting a strong turbulent dissipation.

\begin{figure}
    \includegraphics[width=\linewidth]{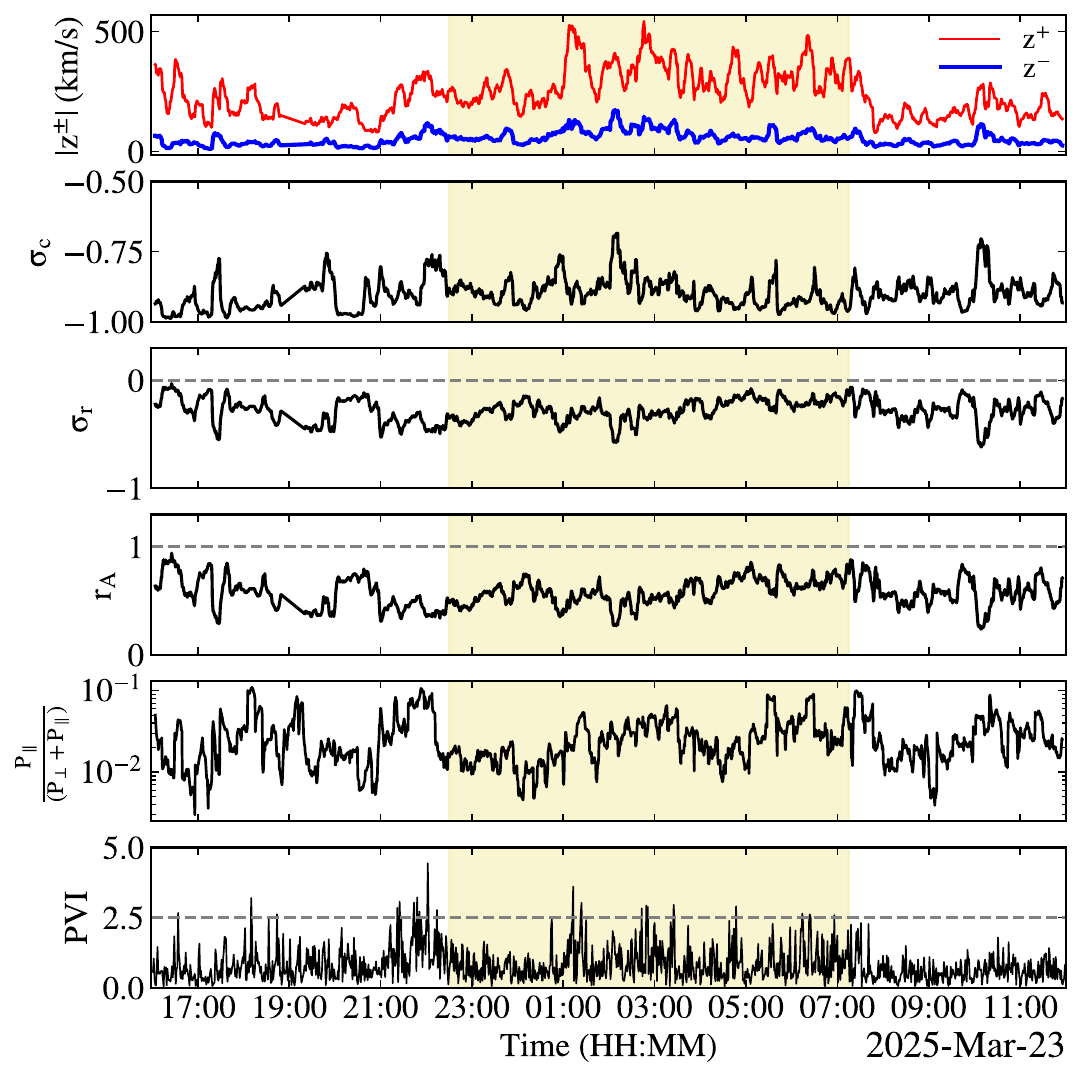}
    \caption{From top to bottom: time series of the Els\"asser variables $z^{+}$ (red) and $z^{-}$ (blue), variance anisotropy $\langle \delta B_{\parallel}^{2}\rangle/\left(\langle \delta B_{\parallel}^{2}\rangle+\langle \delta B_{\perp}^{2}\rangle\right)$, Alfv\'en ratio $r_A$, normalized cross helicity $\sigma_c$, and normalized residual energy $\sigma_r$, and partial variance of increment (PVI). The shaded region denotes the interval of interest. Dashed horizontal lines indicate reference values: 
    $r_A = 1$ (kinetic--magnetic equipartition), and $\sigma_c = 0$, $\sigma_r = 0$ (balanced turbulence conditions).\label{fig:alfvenic}}
\end{figure}

We begin by showing the time series profiles of several turbulence quantities during the crossing of the fast sub-Alfv\'enic interval in Fig.~\ref{fig:alfvenic}. Turbulence properties are estimated for every 10-minute moving windows. The correlation time near PSP perihelia is about $200\,$s~\citep{Parashar2020ApJS}. The Elsasser variables are defined as
$\mathbf{z}^{\pm} = \mathbf{v} \pm {\mathbf{b}}$,
 where $\mathbf{b}$ is the magnetic fluctuation expressed in Alfv\'en speed units. From these variables we compute the normalized cross helicity
\begin{eqnarray}
\sigma_c = \frac{(E^+ - E^-)}{(E^+ + E^-)},\label{eq:sc}
\end{eqnarray}
the normalized residual energy
\begin{eqnarray}
\sigma_r = \frac{(E_v - E_b)}{(E_v + E_b)},
\end{eqnarray}, and the Alfv\'en ratio
\begin{eqnarray}
    r_A = \frac{E_v}{E_b} \label{eq:ra},
\end{eqnarray} 
where $E^\pm, E_v,$ and $E_b$ denote Elsasser, kinetic, and magnetic fluctuation energies. 

To explore the anisotropic nature of the sample, we estimate the variance anisotropy from the magnetic field data. To compute the 
energies in fluctuations in parallel and perpendicular polarizations, 
we transform the magnetic field from the RTN coordinate system to the mean-field coordinate system. Here, 
the mean magnetic field defines the parallel direction and the two remaining orthogonal directions define the perpendicular plane. We then estimate the fraction of parallel fluctuation power ($P_{\parallel} = \langle \delta B_{\parallel}^{2}\rangle$) relative to the total power ($P_{\parallel} + P_{\perp} = \langle \delta B_{\parallel}^{2}\rangle+\langle \delta B_{\perp}^{2}\rangle$).

Figure~\ref{fig:alfvenic} shows the turbulence characteristics of this interval. The plots in this figure reveal some notable properties. Although the observation is made at low latitude (as are all orbits of PSP), the turbulence properties are consistent with the characteristic of PCH-l fast wind. The top panel shows that the Els\"asser amplitudes $z^+$ and $z^-$ are rather dissimilar, with the outward propagating $z^+$ being significantly greater within the fast wind interval. As a result, the normalized cross helicity remains close to  -0.75 to -1.0, indicative of highly Alfv\'enic outward propagating fluctuations. 
Consistently, the Alfv\'en ratio is 
less than unity, indicating dominance of magnetic fluctuations over velocity fluctuations, and so 
the residual energy remains negative. 
The parallel fluctuation power constitutes less than 10\% of the total for most of the fast wind interval. Therefore, the fluctuations are strongly polarized in the directions transverse
to the mean magnetic field. 
All of these parameters suggest that this interval is a case of highly ``Alfv\'enic", highly transverse,
fast wind originating in the middle or lower corona, but here appearing at low heliographic latitude. 

The bottom panel of Figure~\ref{fig:alfvenic} shows the partial variance of increment (PVI) calculated from the magnetic field data using a time lag of 1 minute and a 4~hour averaging window. The PVI values show several peaks occasionally with values larger than $2.5$. This indicates presence of strong current sheets in the selected interval~\citep{Greco2017SSR}.

\begin{figure}
    \includegraphics[width=\linewidth]{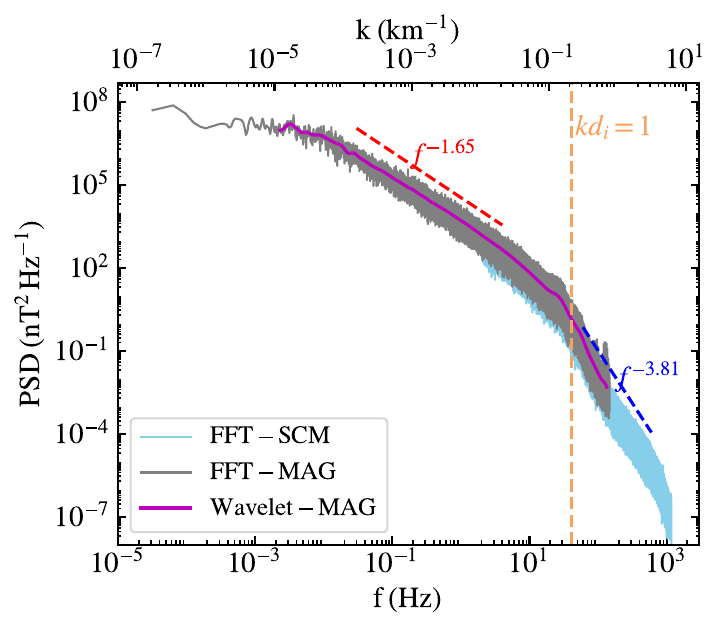}
    \caption{Power Spectral Density (PSD) of the trace of magnetic field fluctuations. 
    The PSD obtained using the Fourier transform and wavelet method are shown for comparison. The red dashed lines represent the power-law fits in the inertial and kinetic ranges. The vertical dashed line denotes the ion inertial length $k\, d_i = 1$.}\label{fig:spec}
\end{figure}

Next, we compute the spectral properties of the magnetic field fluctuations. We convert temporal frequencies ($f$) to wavenumber ($k$) using the modified Taylor hypothesis (similar to \citealt{zank2024characterization}) that accounts for Alfv\'en wave propagation, which is important in near Sun and sub-Alfvénic solar wind intervals where $V_A$ exceeds the solar-wind speed \citep{klein2014violation, zank2022turbulence}. The effective propagation speeds of outward and inward propagating fluctuations are $\mathbf{V}_{\rm eff}^{\pm}=\mathbf{U}_{\rm sc}\pm V_A\,\mathbf{cos \psi}$ \citep{GoldsteinEA86-prop, zhao2022turbulent, zank2022turbulence, zank2024characterization}, where $\mathbf{U_{\rm sc}}$ is the solar wind speed in the spacecraft frame, with $\mathbf{U}=\mathbf{U}_{\rm sc}-\mathbf{V}_{\rm psp}$, where $\mathbf{V}_{\rm psp}$ is the spacecraft velocity. $\psi$ is the angle between the mean magnetic field and flow velocity. 

Fig.~\ref{fig:spec} shows power spectral densities of the magnetic field computed using Fourier transforms (FFT) \citep{welch1967} and wavelet transforms \citep{torrence1998, podesta2009dependence}. Both FFT and wavelet methods produce consistent spectral estimates over the range of frequencies sampled. The inertial range exhibits a power-law scaling close to the Kolmogorov $-5/3$ value, while a clear spectral break is observed near the proton kinetic scale. The break frequency lies near the proton inertial length $(k\,d_i = 1)$, indicated by the dashed vertical line. Beyond the break frequency, the spectrum steepens significantly at higher frequencies, indicating the transition to ion kinetic scales and onset of dissipative processes. The steep powerlaw behavior continues to higher frequencies, as shown by the SCM data. However, we note that only two components of the magnetic field vector are available from SCM while all three components are used for calculating magnetic spectrum from MAG.

 \begin{figure}
    \centering
    \includegraphics[width=\linewidth]{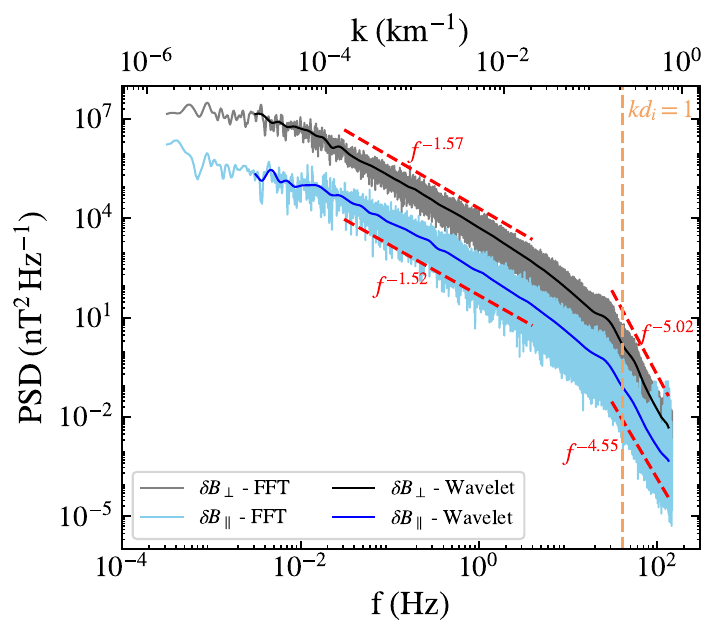}
    \caption{PSD of magnetic field fluctuations decomposed into perpendicular ($\delta B_{\perp}$) and parallel ($\delta B_{\parallel}$) components with respect to the local mean magnetic field (top panel). Light-colored lines represent the PSDs obtained from Fourier Transform and solid lines represent those using wavelet transform.}\label{fig:anisotropy}
\end{figure}

To further explore the scale dependent nature of the anisotropy in magnetic fluctuations in this sample, we plot the components separately in spectral space. Figure~\ref{fig:anisotropy} decomposes magnetic fluctuation
power 
into components perpendicular and parallel to the local mean magnetic field. The perpendicular component dominates throughout most of the inertial range, indicating predominantly transverse fluctuations, while the parallel component becomes relatively more important approaching kinetic scales. The normalized parallel power remains below equipartition over much of the measured range, suggesting a mainly 2D type turbulence for most part.


 \begin{figure}
    \includegraphics[width=\linewidth]{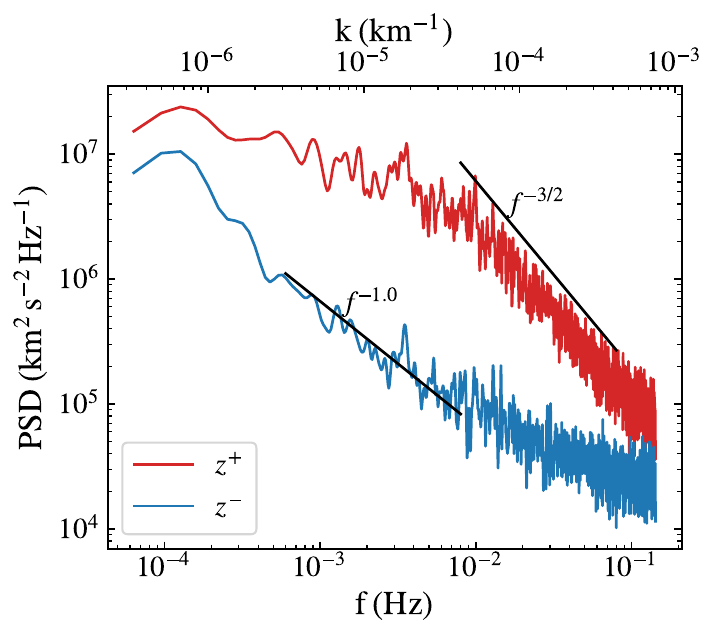}
    \caption{Variation of the Els\"asser amplitude across scale. The upper panel shows the PSD of $z^{+}$ (gray: FFT, black: wavelet) and $z^{-}$ (light blue: FFT, blue: wavelet).
}\label{fig:zpm}
\end{figure}

Figure~\ref{fig:zpm} show the spectra of outward and inward propagating modes. Solid black lines represent the reference slopes. The fitted spectral slopes in the inertial range for $z^+$ and $z^-$ are -1.42 and -0.71, respectively.
We note that the 
spetcral slopes of the $+$ and $-$ Els\"asser fields
are are not separately constant  through the identified inertial range (see Fig \ref{fig:spec})
nor is the ratio of the two spectra at a given frequency
constant in that range. 
This corresponds to the 
statement that the 
frequency dependent normalized cross helicity is not, in general,
constant in frequency 
(wavenumber)
over the inertial range.
Such complex behavior in scale, departing from pure powerlaw ``scaling" 
has been seen in analytical MHD closures \citep{GrappinEA83}
and in MHD simulations
\citep{PouquetEA86}. 
Variations of this behavior 
have also frequently 
been observed 
in solar wind turbulence
\citep{MattGold82a,Tu1995SSR}.
The concave-upward shape of the smaller Els\"asser spectrum
(Fig.\ref{fig:zpm}), trending toward equality with the larger of the two spectra
for frequencies approaching the dissipation scales, is also a familiar feature, and may be due to the effect of noise in the velocity measurements at higher frequencies. 
The variability of the Els\"asser spectra is 
not fully explained theoretically as far as we are aware, and so far appears to be an indicator of the lack of universality in MHD and plasma turbulence \citep{LeeEA10,WanEA12-jfm}. Nevertheless, the Elsasser slopes in Fig.~\ref{fig:zpm} appears to be roughly consistent with the theory developed by \cite{zank2022turbulence}.


\begin{figure}
    \centering
    \includegraphics[width=\linewidth]{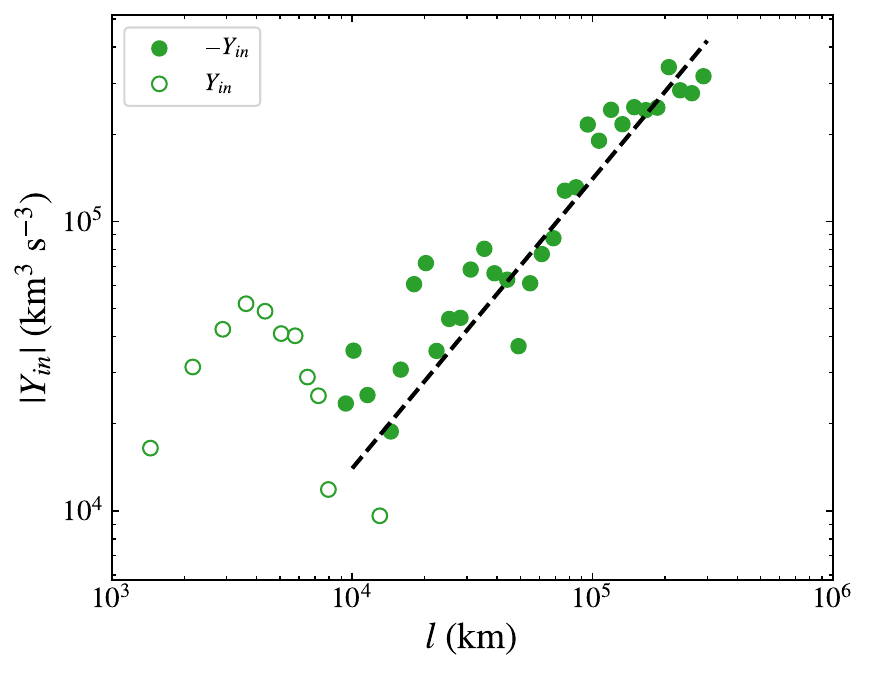}
    \caption{Absolute value of the incompressible Yaglom flux, $|Y_{\mathrm{in}}|$ as a function of spatial lag $\ell$. Negative (filled circles) and positive (open circles) values of $Y_{\mathrm{in}}$ correspond to forward and reverse energy transfer, respectively. The dashed black line denotes the reference linear scaling.\label{fig:yin}
}
\end{figure}

Finally, we present an estimation of the energy cascade rate using
the Kolmogorov-Yaglom law. Here, we use the extension of the isotropic third-order law in MHD~\citep{Politano1998GRL,Politano1998PRE},
\begin{eqnarray}
Y^{\pm}(\ell) = - \frac{4}{3} \epsilon^{\pm} \ell \label{eq:3ord},
\end{eqnarray}
where, $Y^{\pm}$ are calculated from the third-order mixed structure functions of the Els\"asser variables, $Y^{\pm}(\ell)=\left\langle
\left|\Delta \mathbf{z}^{\pm}(\ell)\right|^{2}
\Delta z_{\parallel}^{\mp}(\ell)
\right\rangle,$
where $\Delta \mathbf{z}^{\pm}(\ell)
=
\mathbf{z}^{\pm}(\mathbf{x}+\ell)
-
\mathbf{z}^{\pm}(\mathbf{x})$. The angle brackets denote an ensemble average over all increments separated by the spatial lag $\ell$. The total incompressible Yaglom flux is then defined as $Y_{\rm in}(\ell)=(Y^{+}(\ell)+Y^{-}(\ell)){2}.$

The quantities, $\epsilon^{\pm}$ in equation~(\ref{eq:3ord}), represent the mean decay rate of the respective Els\"asser energies (per units mass): $\epsilon^{\pm} = \mathrm{d} {(Z^{\pm})}^{2} / \mathrm{d}t$; where $Z^{\pm}$ are the root-mean-square fluctuation values of the Elsasser fields. The total energy decay rate can then be calculated as $\epsilon = (\epsilon^{+} + \epsilon^{-})/2$. For single-spacecraft observations, the structure functions are computed for different temporal lags $\tau$. Like bore, we then use the modified Taylor's ``frozen-in'' hypothesis~\citep{Taylor1938PRSLA} to interpret the temporal lags as spatial lags.

Figure~\ref{fig:yin} plots the incompressible Yaglom flux as a function of spatial lags. Negative (filled circles) and positive (open circles) values correspond to forward and inverse energy transfer, respectively. The negative fluxes dominate over most spatial scales. They are broadly consistent with the expected linear scaling, $|Y_{\rm in}| \propto l$, over the inertial range, in agreement with the third-order turbulence phenomenology (Eq.~\ref{eq:3ord}). Although inverse fluxes are also present, particularly at smaller spatial scales, they occur less frequently. The estimated turbulent energy cascade rates span approximately
$10^{7}\ \mathrm{J\,kg^{-1}\,s^{-1}}$. The outward Els\"asser
cascade rate, is typically of the order of
$10^{6}$--$10^{7}\ \mathrm{J\,kg^{-1}\,s^{-1}}$, whereas the inward
cascade rate, is generally smaller, ranging from
about $10^{4}$ to $10^{6}\ \mathrm{J\,kg^{-1}\,s^{-1}}$. This indicates
that the outward-propagating fluctuations dominate the turbulent energy
transfer throughout the inertial range.

\section{Discussion}\label{sec:disc}
The present work reports the first observation of a PCH-l fast solar wind stream in the sub-Alfv\'enic low solar corona, at the low heliographic latitudes probed so far by PSP.
The fast wind emerging from coronal holes is generally associated with strongly outward-propagating Alfv\'enic fluctuations~\citep{Neugebaur1966JGR_average}, as seen here. 
It is significant that 
the interval studied here exhibits some properties consistent with fully developed turbulence, suggesting ongoing strong MHD turbulence deeper in the corona. The Alfv\'enicity parameters, shown in Fig.~\ref{fig:alfvenic}, are similar to those reported by \cite{zhao2022turbulent} for the first extended sub-Alfv\'enic interval observed by PSP. 

The magnetic field spectrum shows a Kolmogorov powerlaw slope in the inertial range. In the ion-kinetic range a steep $\approx -4$ spectrum is observed. The magnetic spectral properties are broadly consistent with the earlier sub-Alfv\'enic observations of \cite{zank2022turbulence}. They reported inertial-range spectral indices of approximately -1.52 and -1.48 for transverse and parallel magnetic fluctuations, respectively, with the transverse fluctuations dominating the magnetic variance. In the present interval, the corresponding perpendicular and parallel magnetic spectra have slopes of approximately -1.57 and -1.52, while the trace magnetic spectrum follows a somewhat steeper -1.65 scaling, close to the Kolmogorov -5/3 value.

Both \cite{zank2022turbulence} and \cite{zhao2022turbulent} obtained inertial-range spectral indices of approximately $-1.5$ and $-1.4$ for the dominant $z^+$ and minority $z^-$ Elsasser modes, respectively. In the present interval, we obtain corresponding slopes of $-1.42$ and $-0.71$. Thus, the dominant $z^+$ spectrum is similar to the previous sub-Alfv\'enic observation and remains close to the Kraichnan $-3/2$ scaling. However, the minority $z^-$ spectrum is somewhat flatter in the present interval. This may indicate the need for a new theoretical framework for this kind of sub-Alfv\'enic wind samples, but more samples will be required to reach a statistically robust conclusion.

The fluctuations here are highly anisotropic, dominated by transverse nearly 2D modes, largely consistent with previous sub-Alfv\'enic studies \citep[e.g.,][]{Bandyopadhyay2022ApJL_subAlfven, Adhikari2026ApJ_sub-alfvenic}.  

The average normalized turbulence amplitude in the present interval is $\delta B /B \approx 0.22$, comparable to the mean values of $0.18$ and $0.22$ reported for the previous sub-Alfv\'enic streams by \cite{Bandyopadhyay2022ApJL_subAlfven, Adhikari2026ApJ_sub-alfvenic}. However, the absolute turbulence amplitude $\delta B$ has a mean value of $\approx 257$ nT, which is much higher than the previous reported values $\sim 100$ nT by \cite{Bandyopadhyay2022ApJL_subAlfven, Adhikari2026ApJ_sub-alfvenic}.

The mixed, third-order structure function
\citep{Politano1998GRL,Politano1998PRE}, constructed from the Els\"asser variables, exhibit a linear scaling through the inertial range of scales, giving a 
local solar wind heating rate of $\gtrsim 10^6 - 10^7\,{\mathrm{W\,kg^{-1}}}$.
This is a higher cascade rate than values observed at 1 au ($\sim 10^3 \,{\mathrm{W\,kg^{-1}}}$) \citep{stawarz2009turbulent, Gautam2024ApJ_cascade}
and also somewhat higher than the values of a few $10^6 \,{\mathrm{W\,kg^{-1}}}$
estimated from prior measurements using PSP observations in sub-Alfv\'enic streams~\citep{zhao2022turbulent}.

Thus, this unique fast PCH-like sub-Alfv\'enic stream exhibits some highly-Alfv\'enic and strongly anisotropic turbulence characteristics very similar to those previously observed in sub-Alfv\'enic wind, but the spectral slope, the level of fluctuations, and cascade rates are distinct from the previous samples.

This observation suggests that the PCH-l fast solar wind, emerging from coronal holes, may already have sufficient time to develop non-linear turbulent fluctuations transverse to the mean field. Several possible mechanisms may contribute to the development of turbulence in these regions, including enhanced wave reflection in the low corona, nonlinear turbulent interactions, or interactions with neighboring streams and magnetic structures~\citep{Matthaeus1999ApJ, Ruffolo2020ApJ_switchback}. 
Even though the outward fluctuations 
are dominant energetically, the 
minority inward Els\"asser energy is apparently great enough to support clear indications of turbulence, 
including
a robust cascade rate obtained from the third order law \citep{Sorriso-Valvo2007PRL,Bandyopadhyay2020ApJS_cascade}. This may indicate that wave reflection and instability processes are already active below the Alfv\'en surface,
as would be necessary for a wind accelerated by reflection driven turbulent heating \citep{Matthaeus1999ApJ}.

So far, the current interval provides the only observation of persistent fast solar wind in the sub-Alfv\'enic corona. Further statistical analysis of similar PSP intervals is needed to determine whether PCH-l fast Alfv\'enic wind at low latitudes
represents a distinct solar wind class or a transitional evolutionary state associated with specific coronal source regions and magnetic geometries.

\section{Acknowledgements}
We are deeply indebted to everyone who helped make the Parker Solar Probe (PSP) mission possible. Parker Solar Probe was designed, built, and is now operated by the Johns Hopkins Applied Physics Laboratory as part of NASA’s Living with a Star (LWS) program (contract NNN06AA01C). Support from the LWS management and
technical team has played a critical role in the success of the
Parker Solar Probe mission. The data used in this paper are
publicly available via the NASA Space Physics Data Facility
(https://spdf.gsfc.nasa.gov/). The QTN-based electron density data are available at the CNES Data Archive for CDPP (https://cdpp-archive.cnes.fr/user/cdpp/modules/1778). We thank CNES (Centre National d'Etudes Spatiales) and CNRS (Centre National de la Recherche Scientifique) for their support and the CDPP (Centre de Donn\'ees de la Physique des Plasmas, CNES, Toulouse, France) for the data distribution. This research was partially supported by by the U.S. National Science Foundation FDSS award AGS-2347952 at the University of Delaware and in part by the PSP/IS\(\odot\)IS project through subcontract SUB0000165 from Princeton and PUNCH project through subcontract N99054DS from NASA/SWRI to the University of Delaware. S.T.B and Y.J.R. were partially supported by Parker Solar Probe project through the SAO/SWEAP subcontract 975569.

 \newcommand{\BIBand} {and} 
\newcommand{\boldVol}[1] {\textbf{#1}} 
\providecommand{\SortNoop}[1]{} 
\providecommand{\sortnoop}[1]{} 
\newcommand{\stereo} {\emph{{S}{T}{E}{R}{E}{O}}} 
\newcommand{\au} {{A}{U}\ } 
\newcommand{\AU} {{A}{U}\ } 
\newcommand{\MHD} {{M}{H}{D}\ } 
\newcommand{\mhd} {{M}{H}{D}\ } 
\newcommand{\RMHD} {{R}{M}{H}{D}\ } 
\newcommand{\rmhd} {{R}{M}{H}{D}\ } 
\newcommand{\wkb} {{W}{K}{B}\ } 
\newcommand{\alfven} {{A}lfv{\'e}n\ } 
\newcommand{\alfvenic} {{A}lfv{\'e}nic\ } 
\newcommand{\Alfven} {{A}lfv{\'e}n\ } 
\newcommand{\Alfvenic} {{A}lfv{\'e}nic\ }

\end{document}